\documentclass[12pt]{article}
\def\beqar {\begin{eqnarray}}
\def\eeqar {\end{eqnarray}}
\def\beq {\begin{equation}}
\def\eeq {\end{equation}}

\def\half {{\textstyle{1\over 2}}}

\def\del {{\partial}}
\def\bdel{\bar{\partial}}

\def\bz {\bar{z}}

\def\no2 {{\textstyle{n\over 2}}}

\begin{document}
\begin{titlepage}
\null\vspace{-62pt}

\pagestyle{empty}
\begin{center}
 \rightline{CCNY-HEP-04/7} \rightline{August 2004}

\vspace{1.0truein}
{\Large\bf  Nonsingular 4d-flat branes in six-dimensional }\\
\vskip .2in
{\Large\bf supergravities}\\
\vskip .1in \vspace{.4in}
V.~P.~NAIR\\
\vskip .1in
Physics Department\\
City College of the CUNY\\
New York, NY 10031\\
E-mail: vpn@sci.ccny.cuny.edu\\
\vskip .4in

S.~RANDJBAR-DAEMI \vskip .1in
Abdus Salam International Centre for Theoretical Physics\\
Trieste, Italy\\
E-mail: seif@ictp.trieste.it\\
\end{center}
\vskip .3in \centerline{\bf Abstract} \vskip .1in

We show that six-dimensional supergravity models admit nonsingular
solutions in the presence of flat three-brane sources with
positive tensions. The models studied in this paper are nonlinear
sigma models with the target spaces of the scalar fields being
noncompact manifolds. For the particular solutions of the scalar
field equations which we consider, only two brane sources are
possible which are positioned at those points where the scalar
field densities diverge, without creating a divergence in the
Ricci scalar or the total energy. These solutions are locally
invariant under ${1\over 2}$ of D=6 supersymmetries, which,
however, do not integrate to global Killing spinors. Other branes
can be introduced by hand by allowing for local deficit angles in
the transverse space without generating any kind of curvature
singularities.

\end{titlepage}

\hoffset=0in
\newpage
\pagestyle{plain} \setcounter{page}{2}
\newpage
\section{Introducion}

The idea of solving the cosmological constant problem by regarding
our universe as a brane in more than four dimensions opens up new
interesting possibilities for attacking this fundamental question
\cite{Rubakov:1983bz}. In the context of the brane-world models
the uniform part of the cosmic energy is indistinguishable from
the brane tension.  The nonzero tension of a  physical
three-brane, on which the particles of the standard model are
localized, may be compensated by a higher dimensional curvature or
a bulk cosmological constant, without generating curvature in the
brane.

In a recent work it was shown that in a nonsupersymmetric
six-dimensional model of gravity coupled to a sigma model targeted
on any K\"ahler  target space one can obtain nonsingular solutions
in the presence of flat three-branes with positive tensions
\cite{Randjbar-Daemi:2004ni}. The branes in that paper could be
viewed as vortices distributed on a two-dimensional compact
manifold with the topology of a two-sphere $S^2$.  The vorticity of a brane
can be thought of as an Aharonov-Bohm phase acquired by  a scalar
field when a complete rotation around the brane is made. There is
a single relation between the tensions of the three-branes and
their vorticities and it was speculated that the time-variation of
the vorticity of each brane may account for the adjustment of the
four-dimensional cosmological constant.  To obtain such a
solution, however,  it was necessary to assume that the
six-dimensional cosmological constant was zero.  It was indeed
shown subsequently in \cite{Lee:2004vn} that a nonzero
six-dimensional cosmological term disturbs the flatness of the
three-branes.

In this paper we shall find solutions  in six-dimensional
supergravity models where a bulk cosmological constant is
forbidden by supersymmetry. We shall show that the tensions of the
branes can be arbitrary  and positive subject to a single linear
relation of the type stated in \cite{Randjbar-Daemi:2004ni} . The
solution for the scalar field will be implicit and more involved
than the simple mappings given in \cite{Kehagias:2004fb,
Randjbar-Daemi:2004ni, Lee:2004vn}. Flat three-brane solutions
with a singularity at the boundary of the transverse space are
known to exist in $D=6$ supergravities with scalars described by a
nonlinear sigma model \cite{Kehagias:2004fb, DRS}. This
singularity is, unfortunately, located at a finite proper distance
from the brane or from any other point  on the transverse space.
Further, although the Killing spinor equations have local
solutions far away from the brane, they do not integrate to a
single-valued global solution \cite{Kehagias:2004fb}.

The aim of this paper is to present nonsingular solutions in a
supergravity  model, where like in \cite{Kehagias:2004fb}, apart
from gravity, the only other active fields will be a set of scalar
fields with dynamics governed by a nonlinear sigma model
Lagrangian targeted on some noncompact hyperbolic or quaternionic
manifold.  Such scalar fields exist in all supergravity models in
six dimensions. Our construction will lead to a smooth transverse
space with no boundaries and with an Euler number of +2. Like the
solution in \cite{Kehagias:2004fb, DRS} our solution in this paper
will also be only locally invariant under 1/2 of D=6
supersymmetries. These local supersymmetries, however, do not
integrate to globally single valued Killing spinors.

Similar attempts to solve the cosmological constant problem have
been made in the past by invoking the magnetic  monopole
compactifiction  in a six dimensional theory of gravity coupled to
Maxwell field and in the presence of a six-dimensional
cosmological constant \cite{Carroll:2003db, Navarro:2003vw}.  The
Kaluza-Klein solution of this system needs a very special tuning
of six-dimensional cosmological term versus the radius of the
internal  $S^2$ in order to have a flat four-dimensional part
\cite{Randjbar-Daemi:1982hi}.  A similar  tuning is also necessary
in the $(1,0)$ six-dimensional gauged supergravities of the type
constructed in \cite{Salam:1984cj, Randjbar-Daemi:1985wc},
although in this case the $Minkowski_4\times S^2$ turns out to be
the unique maximally symmetric solution which also preserves
${1\over 2}$ of the supersymmetries \cite{Gibbons:2003di}. These
theories also have three-brane solutions but with negative
tensions and/or singular transverse spaces
\cite{Aghababaie:2003wz, Gibbons:2003di,
Aghababaie:2003ar}.\footnote{Some of these issues will be further
discussed in a forthcoming paper by C.Burgess {\it et al}.}

The solution of the scalar field equation in this paper is akin to
the one studied in the context of the string comic string in
\cite{Greene:1989ya}. There, the complex scalar field is the
modulus of a $T^2$, therefore, modular invariance plays a very
important role. In our case
  the restriction imposed by modular invariance
is needed for the single valuedness of the space time metric. For
this reason our solution for the metric will be somewhat different
and, unlike the metric in \cite{Greene:1989ya}, the singularities
in our metric will be smoothed out. This smoothing out necessarily
introduces delta function type terms in the Einstein equations
whose natural interpretation is the presence of three-branes.
Demanding regularity at infinity and positivity of the tensions of
the branes will impose more restrictions on our solution to the
scalar field equation compared to the solution in
\cite{Greene:1989ya}. However, unlike the monopole solutions of
\cite{Carroll:2003db, Navarro:2003vw}, which require  tuning of
the parameters to achieve flatness,we shall need a single linear
relation on the tensions. The single relationship between the
tensions is dictated by the requirement that the geometry becomes
spherical away from the branes, as in
\cite{Randjbar-Daemi:2004ni}. This reference also introduced the
notion of vorticity for the brane solutions, but, because of the
complexity of the scalar field ansatz, it is not clear to us how
we can extend this idea to the present case in a straightforward
way.

The plan of this paper is as follows: In section 2 we present the
solution and show its regularity. In section 3 we examine the
conditions for unbroken supersymmetries. Section 4 is dedicated to
a brief summary.

\section{The Solutions}

Turning to the details of the solutions, consider a $D=6$
supergravity with scalars in the hyper or tensor multiplets. Such
scalars cover a noncompact hyperbolic or quaternioic manifold of
the type $SO(n,1)/SO(n)$ for the scalars in the tensor multiplet
or $Sp(n,1)/Sp(n)\times Sp(1)$ for the scalars in the
hypermultiplet. (In this latter case some other quaternionic
K\"ahler manifolds are also possible \cite{Nishino:1986dc}. Our
discussion will be applicable to all of them.) We shall set all of
the scalar fields to zero apart from a single complex one which we
shall denote by $\varphi$.  The metric in the two-dimensional
subspace of the scalar manifold is given by
\beq ds^2= \frac
{d\varphi ~d\bar\varphi}{(1-\vert \varphi\vert^2)^2}
\label{metric} \eeq
A simple holomorphic change of coordinates
$\varphi = ( \tau - i)/(\tau +i)$ will then bring it to the form
\beq
ds^2 = - \frac{d\tau d \bar \tau}{(\tau-\bar\tau)^2}\label{1}
\eeq
We shall look for solutions of the Einstein's equations of
the form \beq ds^2= \eta_{\mu\nu}dx^\mu dx^\nu + e^{\phi(z, \bar
z)} dz d\bar z \label{2} \eeq where $\mu, \nu = 0,1,2,3$ and $(z,
\bar z)$ is a local complex coordinate in the extra two
dimensions. In this coordinate system the relevant field equations
reduce to \beq \del \bdel \tau + 2~ { \del \tau \bdel \tau \over
{{\bar\tau}-\tau}} =0 \label{3} \eeq for the scalar field and \beq
\del \bdel \phi =  { \del \tau \bdel {\bar\tau} \over
{{(\bar\tau}-\tau)^2}} - \frac{1}{2M^4} \sum T_i \delta
^{(2)}(z-z_i) \label{4} \eeq for the metric function of the
transverse two-space.  In the last equation, $M$ is the
six-dimensional Planck mass and $T_i$ are the tensions of the
brane sources. Note that, unlike in the case of the
nonsupersymmettric model, in the present case, the $D=6$
supersymmetry does not allow for an independent coupling constant
for the sigma model. Apart from the brane source terms the form of
the field equations are completely fixed by supersymmetry. We also
note that the field equations are invariant under an $SL(2,R)$
group acting on $\tau$ by fractional linear transformations, viz.,
\beq \tau'(z) = \frac{a\tau(z) +b}{c\tau(z)+d}\label{5} \eeq where
$a,b,c,d$ are real numbers satisfying $ad-bc=1$.

We first consider the scalar field equation. Clearly any function
of $z$ which is independent of $\bar z$ solves that equation. In
particular, we take the matter field configuration to be given by
\beq J(\tau(z) )=  {z-a \over z-b} \label{6} \eeq where $J(\tau )$
is the modular function known as the absolute invariant
\cite{witt}. By solving this for $\tau$ in terms of $z$, we get
the matter field configuration. Since $J$ is invariant under
modular transformations $SL(2,Z)$ of $\tau$, the inverse will be
multivalued unless we restrict to the fundamental domain $F_0$ of
$\tau$.  Writing $\tau= \tau_1 +i\tau_2$, this domain is defined
by \beq F_0 = \{ \tau ~\vert~ \tau_2 >0, ~ \vert \tau \vert \geq
1, ~ -\half \leq \tau_1 < \half , ~~ \tau_1 \leq 0 ~ {\rm for}
\vert \tau \vert =1 \} \label{7} \eeq In this domain, there is a
unique inverse. We will define our solution as the configuration
$\tau (z)$ where $\tau$ takes values in the fundamental region.
Other regions correspond to other solutions which are degenerate
with this;  classically they can be treated as separate and
different solutions. Also, since the equations of motion are
invariant under a global $SL(2, R)$ action, we can obtain new
solutions by acting with this group on any given solution.

Note that, as far as the $\tau$ field equations are concerned,
there is an infinite class of solutions of the type suggested
here. In principle, we could consider an implicit equation of the
form $J(\tau (z))= f(z)$, where $f$ is any holomorphic function of
$z$. Here we have taken $f$ to be a simple rational function which
approaches unity as $z$ goes to infinity. We shall see later that
the requirement of the positivity of the tensions of the branes
and regularity of the metric as $\vert z\vert$ goes to infinity
exclude every other possibility for $f(z)$ apart from the simple
one given above.

The ansatz for the metric is given by
\beqar
 e^\phi &=& \gamma^2~\left[
\tau_2 \eta^2 {\bar \eta}^2{1\over \vert z-b\vert^{2\over12}}\right]
~{\vert z-a\vert^{-2\alpha} } { \vert z-b\vert^{-2\beta}}\nonumber\\
&&\hskip.2in \times
\exp \left[ \int d^2 z' G(z,z')  { \bdel \del \tau(z') \over
{{(\bar\tau}-\tau)(z')}}\right] \label{8}
\eeqar
where $\alpha,
\beta,\gamma$ are constants and the Green function $G(z,z')$
satisfies \beq \bdel \del G(z,z')= \delta^{(2)}(z-z') \label{9}
\eeq The exponential involving the Green function tends to a
constant whenever $z$ approaches $a$ or $b$, which, as we shall
see, will be identified with the positions of the three-branes,
and it simply vanishes at every other point.
$\eta (\tau )$ is the Dedekind $\eta$-function. The first factor
$ \tau_2 \eta^2 {\bar\eta}^2 \vert z-b\vert^{-{2\over 12}}$ is needed for modular invariance. Since for
$z\rightarrow b$ the $\eta$ function behaves like $(z-b)^{1/24}$, the factor
$ \vert z-b\vert^{-{2\over 12}}$ ensures that the metric does not vanish as $z$ approaches $b$.

Notice that, apart from $\tau_2$ and the exponential factor
involving $G(z,z')$,  $e^\phi$  is of the form $g(z) {\bar g}(\bz
)$, where $g(z)$ is holomorphic in $z$. As a result, in the
calculation of $\del \bdel \phi$, the function $g$ only
contributes $\delta$-functions, which can be identified as the
brane contribution to the energy-momentum tensor. Since the factor
$\eta / (z-b)^{1/12}$ has no singularity at $z =b$,  the contribution
due to the brane at $z=b$ comes entirely from the factor
$\vert z-b\vert^{-2\beta}$. The contribution
due to $\tau_2 \exp [\int d^2 z' G(z,z')  { \bdel \del \tau(z') /
{{(\bar\tau}-\tau)(z')}}] $ will exactly match the field part of
the energy-momentum tensor. Also the constant prefactor $\gamma$
in $e^\phi$, which measures the size of the transverse space,  is
undetermined by the field equations and is thus a modulus.

We will now turn to an analysis of the scalar  field, metric and
curvature at various special points, which will yield constraints
on the exponents $\alpha$ and $\beta$. But before doing so, let us
note that the tensions at these points are simply given by \beq
T_a= 2\pi M^4\alpha\quad\quad\quad\quad T_b= 2\pi M^4 \beta
\label{10} \eeq This result follows from the substitution of our
ansatz for $e^\phi$ in the Einstein equation and then equating the
coefficients of the delta functions at $z=a$ and $z=b$.

We will now require that the transverse geometry in the vicinity
of the location of the branes should have no curvature
singularities; this will put restrictions on $\alpha$ and $\beta$.
There are three special points to analyze  \cite{witt}. These
points are essentially characterized by the condition of the
vanishing of $J(\tau)$ or $\frac{dJ}{d\tau}$. By analyzing our
equations above, it is easy to see that these are the only points
where there is the danger of curvature singularities.
\begin{enumerate}
\item At $z=a$, $J$ and $\frac{dJ}{d\tau}$  vanish and $\tau =
\omega \equiv \exp (2\pi i /3)$. Near this point, we can expand
$\tau$ as \beq \tau \approx \omega + c ~(z-a)^{1\over 3}
\label{10a} \eeq where $c$ is a constant. $\tau_2$ and the
exponential factor in $e^\phi$ are finite at $z=a$,  so that the
metric has the behavior \beq e^\phi \sim {1\over \vert z -a
\vert^{2\alpha }} \label{11} \eeq
The geometry has a conical
defect at $z=a$ . The energy density for matter fields behaves as
\beq {\cal E} \approx {\vert c\vert^2 \over 27} {1\over \vert
z-a\vert^{4\over 3}} \label{12} \eeq The energy density has
singularity; this will be matched exactly by a similar singularity
in $\del \bdel \phi$ so that the Einstein equation will be
satisfied. Further, this is an integrable singularity, so that
there is no pathology in the total energy. In fact, the curvature
is given by \beqar R&=& - e^{-\phi } \left[ -{\vert c\vert^2 \over
27} {1\over \vert z-a\vert^{4\over 3}} -\pi \alpha \delta^{(2)}
(z-a)
\right] \nonumber\\
&\sim&  \vert z-a\vert^{2\alpha}\left[ {\vert c\vert^2 \over 27}
{1\over \vert z-a\vert^{4\over 3}} + \pi \alpha \delta^{(2)} (z-a)
\right]\label{13}\\
&=& e^{-\phi} \left[ {\cal E} + \pi \alpha \delta^{(2)} (z-a)
\right]\nonumber \eeqar We notice that there is a singularity in
the curvature if we do not have the term $\vert
z-a\vert^{2\alpha}$. With this term, we can avoid the singularity
by taking \beq \alpha \geq {2\over 3} \label{14} \eeq In view of
the aforementioned relationship between the tension of the brane
at $z=a$ and the parameter $\alpha$ we thus obtain a lower bound
on $T_a$, namely \beq T_a \geq {4\pi\over 3}M^4 \label{t} \eeq

The behavior of $\tau$ near $z=a$ requires another comment.
Evidently, $(z-a)^{1\over 3}$ is not single-valued as we go around the
point $z=a$; also the value of $\tau$ after such a rotation,
namely, $\tau (\lambda e^{2\pi i})$, $\lambda = z-a$,
is not within the fundamental modular region.
However up to first order in $\lambda$, there is a modular transformation
given by
\beq
\tau (\lambda e^{2\pi i}) = {{\tau (\lambda ) + 1} \over {-\tau (\lambda )}}
\label{modtran}
\eeq
which maps $\tau$ back into the fundamental region. Since the fields are defined
up to such a modular transformation, this does not change the physical fields.
The factor $ \tau_2 \eta^2 {\bar\eta}^2 \vert z-b\vert^{-{2\over 12}}$
in (\ref{8}) is modular invariant, making the metric invariant,
as in \cite{Greene:1989ya}.

\item As $\vert z\vert \rightarrow \infty$, $J \rightarrow 1$, and
$\tau \rightarrow i$. Thus asymptotically, we can expand $\tau
(z)$ as \beq \tau \approx i + c_2 ~ z^{-\half } \label{15} \eeq
It
is then easily seen that the metric behaves as
\beq e^\phi \sim
\vert z\vert^{-2\alpha -2\beta -{2\over12} } \label{16}
\eeq
In order for this
not to lead to deficit angles at infinity, we need $e^\phi$ to
behave as $\vert z\vert^{-4}$. This gives a constraint \beq
\alpha +\beta +{1\over 12} = 2 \label{17} \eeq which translates into a
constraint among the tensions, viz, \beq T_a + T_b = {11\over 6}~ 2\pi M^4,
\label{18} \eeq an identity similar to the constraint derived in
\cite{Randjbar-Daemi:2004ni}. Note that this behavior as $\vert
z\vert$ goes to infinity also insures that the Euler number of our
transverse space is +2.

\item As $z \rightarrow b$, $J\rightarrow \infty$. In this case,
the metric has the form \beq e^\phi \sim \left[ -{1\over 2\pi}
\log \vert z-b\vert \right] ~\vert z-b\vert^{-2\beta} \label{19}
\eeq Without the extra term $\vert z-b\vert^{-2\beta}$, there is a
curvature singularity, since the Ricci scalar is given by \beq R
\approx \vert z-b \vert^{2\beta} {1\over ( -\log\vert z-b\vert )}
\left[ {1\over 4}{1\over \vert z-b\vert^2 \left( \log \vert
z-b\vert \right)^2 } + \pi \beta \delta^{(2)}(z-b) \right]
\label{20} \eeq
\end{enumerate}
Even though the Einstein equation can be satisfied, balancing the
singular term of ${\cal E}$ with the similar term in $e^\phi R$,
the curvature itself will be singular unless we choose \beq \beta
> {1} \label{a} \eeq In terms of the tension $T_b$ this implies
\beq T_b >  2\pi M^4 \label{t'} \eeq Thus, we see that, to
obtain regular geometry not only the tensions should be bounded from
below by positive quantities,  their sum should also be bounded by
$(11/6)\times 2\pi M^4$. It can be verified that are no other points at which
the curvature behaves badly. The three constraints do have a set of
solutions for $\alpha$ and $\beta$.

Let  us now turn to a possible generalization of equation(\ref{6})
to \beq J(\tau(z)) =  \prod_{k=1}^N ~\frac {z-a_k}{z-b_k}
\label{20a} \eeq
In order to avoid singularities whenever $z$ approaches any one of
the $a_k$'s or $b_k$'s we need to insert  a factor $\prod _k \vert
z-{a_k}\vert ^{-2\alpha_k}  \vert z-{b_k}\vert^{-2\beta_k}$
in the ansatz for the metric and
restrict each $\alpha_k$ and $\beta_k$ by \beq \alpha_k \geq \frac
{2}{3}, \quad\quad\quad \beta_k \geq 1 \label{21} \eeq The
requirement  that $e^\phi$ should behave like  $\vert z\vert
^{-4}$
whenever  $\vert z\vert$ goes to infinity will then restrict these
parameters by the linear relation \beq \sum_1 ^N (\alpha_k +
\beta_k) +{1\over 12} = 2 \label{22} \eeq
Clearly we can satisfy all these
constraints only for $N=1$. Presumably the introduction of
vorticity as in \cite{Randjbar-Daemi:2004ni} would allow us more
freedom. However, at this stage it is not clear to us how to do
this a simple physical way.

The only way that we can  keep $\tau (z)$ as a solution of the
simple equation $J(\tau(z))= (z-a)/(z-b)$ and  yet introduce more
than two branes is to modify the metric by inserting  a factor
like  $ \vert z-a\vert ^{-2\alpha}  \vert z-b\vert^{-2\beta}\prod
_k \vert z -{a'_k}\vert ^{-2\alpha_k}  \vert z
-{b'_k}\vert^{-2\beta_k}$, where $a_k's$ and $b_k's$ are different
from $a$ and $b$.  In this case, by requiring that $\alpha'_k$'s
and $\beta'_k$'s are smaller than unity, we can have a solution to
all the constraints.

\section{ Supersymmetry}
Now we turn to the question of unbroken supersymmetries. Having
set all the fields to zero apart from gravity  and a pair of real
scalars which we have assembled into a single complex field
$\tau$,  the only Killing spinor equations we need to examine are
those of gravitino and the scalars. Although our argument is valid
for all ( ungauged)  supergravities in D=6, to be concrete, in this
section, we shall consider hypermatter scalars in ungauged (1,0)
supergravities. Such models are chiral and can be obtained from
the compactification of the heterotic string on $K3$
\cite{Green:1984bx} or the M theory on $K3\times S^1/Z_2$
\cite{Sen:1996tz}. They parameterize a quaternionic manifold of
the form $G/H\times SU(2)$.

The gravitino as well as the hyperino carry spinor indices of
SU(2). Their supersymmetry variation in our bckground is given by
\footnote{Here we closely follow the notations of
\cite{Nishino:1986dc} and \cite{Randjbar-Daemi:2004qr}.}

\beq \delta\psi_m ^r = \partial_m \epsilon^r + \frac{1}{2}\omega
_m  \Gamma_{45} \epsilon^r + \partial_m\phi^\alpha Q_\alpha
^{rs}\epsilon_s \eeq

 \beq
\delta\psi^{\dot r r}= V_\alpha ^{\dot r r}\partial_m \phi^\alpha
\Gamma^m \epsilon_r
\eeq
 In these equations the spinor indices $r, s...$ take values 1 and
 2, the index $m$ is tangent to the transverse space and takes the
 values 4 and 5. $\phi^\alpha$ denote the two nonvanishing
 hyperscalars. $\omega_m$  and $Q$ in the gravitino equation denote
 the connections in the transverse space and the hyperscalar
 manifold respectively. Finally $ V_{\alpha}^{\dot r r}=
  V_{\alpha}^1 (i\sigma^3)^{\dot r
 r} + V_{\alpha}^2 (i\sigma^1)^{\dot r r}$. We have denoted the
 components of the zweibein in the tangent space of the scalar
 manifold by $V_{\alpha}^1$ and $V_{\alpha}^2$. All we need to
 know about them is that their only non vanishing components are
 $V_1 ^1 = V_2 ^2$. Finally we quote, without giving the details,
 that for our background the spin connections $\omega_m$ and $Q$
 are given by

  \beq
  \omega_z = \frac
 {1}{2}\frac{\partial_z
 \tau_1}{\tau_2}+ \partial_\tau \ln \eta \partial_z \tau- \frac{\alpha}{2}\frac{1}{z-a} - \frac{\beta+1/12}{2}\frac{1}{z-b}
 \eeq

 \beq  \partial_z \phi^\alpha Q_\alpha^{rs}=
- \frac{i}{4}\frac{\partial_z
 \tau_1}{\tau_2}(\sigma_2)^{rs}
\eeq

The condition $\delta \psi^{\dot r r}=0$, can be solved provided
we restrict the supersymmetry parameter $\epsilon$ by

 $$[\Gamma_4 (\sigma_3) ^{\dot r r} + \Gamma_5 (\sigma_1) ^{\dot r
 r}]\epsilon_r =0$$

This condition cuts the number of components of $\epsilon$ by
half. Upon imposing this condition the gravitino equation $\delta
\psi_m ^r =0$ then reduces to a first order linear differential
equation for $\epsilon$ as a function of $z$ and $\bar z$ which in
general can be integrated. The solution will, however, be a
mutivalued function of $z$. A full rotation around the branes will
not bring $\epsilon$ back to its original value. It has been
suggested in \cite{Kehagias:2004fb} that by introducing a flat
$U(1)$ connection one can eliminate the multivaluedness by an
Aharonov-Bohm phase. It will be interesting to find a supergravity
model in six dimensions where this idea can be implemented without
disturbing other nice features of our solutions.

\section{Conclusion}
 In summary, we
constructed nonsingular flat brane solutions in ungauged six
dimensional supergravities with a transverse space of Euler number
+2. Since our transverse space has no singularities or boundaries
the value +2 for its Euler number implies that, topologically, it
is a $S^2$. Thus its volume will be finite, although it is not
easy to evaluate it explicitly. The size of the space depends on
an integration constant which is not determined at the classical
level. It is thus a modulus. It will be interesting to see if
temperature effects in a cosmological context can stabilize the
value of this radius. This is known to happen in Kaluza Klein
cosmology in the absence of branes \cite{Randjbar-Daemi:1983jz}.
Aspects of brane cosmology and the question of self tuning of
cosmological constant  in co-dimension two have been studied in
\cite{Cline:2003ak}.

Like the solutions in \cite{Kehagias:2004fb} the solutions
presented in this paper also break ${1\over 2}$ of the
supersymmetries.  The local solution to the Killing spinor
equation cannot, however, be integrated to a single-valued global
solution. It is suggested in \cite{Kehagias:2004fb} that by
coupling the gravitino to a U(1) gauge field and allowing for
multivalued Killing spinors one may recover single-valuedness.
Such U(1) couplings do indeed exist in the gauged six-dimensional
supergravities. However, one difficulty, among others, in
implementing this interesting idea is that in this case the scalar
fields will have a nontrivial potential and their field equations
cannot be solved by holomorphic configurations.\footnote {For an
explicit form of the potential in (1,0) supergravity in six
dimensions see \cite{Randjbar-Daemi:2004qr}.}

Our construction cannot be considered a completely satisfactory
solution to the cosmological constant problem, principally, due to
the linear relation among the tensions that we derived in section
3. The introduction of vorticity along the lines of
\cite{Randjbar-Daemi:2004ni},  may improve our construction,
although we do not have a simple physical way of implementing this
idea in the present context.

\section*{Acknowledgments}
We would like to thank Bobby Acharya for reminding us of the
reference \cite{Greene:1989ya} The work of V.P.N. was supported in
part by the U.S. National Science Foundation grant number
PHY-0244873.

\end{document}